\documentclass[aps,preprint,nofootinbib,floatfix,prd]{revtex4-1}
\usepackage{graphicx,epsfig}
\usepackage{color}
\usepackage{hyperref}

\begin{document}

\title{\bf Glueball Production via Gluonic Penguin B Decays}

\vspace{1cm}

\renewcommand{\thefootnote}{\fnsymbol{footnote}}

\author{
Xiao-Gang He$^{1,2,3}$
\footnote{Email address: {\sf hexg@phys.ntu.edu.tw}}
and Tzu-Chiang Yuan$^{4}$
\footnote{Email address: {\sf tcyuan@phys.sinica.edu.tw}} }
\affiliation
{1.~INPAC, SKLPPC, and Department of Physics,
Shanghai Jiao Tong University, Shanghai 200240, China\\
2.~Physics Division, National~Center~for~Theoretical~Sciences,~and ~Department~of~Physics,~National~Tsing~Hua~University, and Hsinchu, Taiwan\\
3.~Department~of~Physics,~National~Taiwan~University,~Taipei,~Taiwan~10764,~Taiwan\\
4.~Institute~of~Physics,~Academia Sinica, Nangang, Taipei 11529, Taiwan}

\renewcommand{\thefootnote}{\arabic{footnote}}

\date{\today}

\vspace{2cm}

\begin{abstract}
We study glueball $G$ production in 
gluonic penguin decay $B\to G + X_s$, 
using the next-to-leading order $b\to s g^*$ gluonic penguin
interaction and effective couplings of a glueball to two perturbative gluons.
Subsequent decays of a scalar glueball are described by using techniques of effective chiral Lagrangian 
to incorporate the interaction between a glueball and pseudoscalar 
mesons. Mixing effects between the pure glueball with 
other mesons are considered. 
Identifying the $f_0(1710)$ to be a scalar glueball, 
we find that both the top and charm penguin to be important and obtain a sizable
branching ratio for $B\to f_0(1710) + X_s$ of order  $1.3\times 10^{-4}(f/0.07\mbox{GeV}^{-1})^2$,
where the effective coupling strength $f$ is estimated to be $0.07$ GeV$^{-1}$ using experimental data for the 
branching ratio of $f_0(1710) \to K \overline K$ based on chiral Lagrangian estimate. 
An alternative perturbative QCD based estimation of $f$ is
a factor of 20 larger, which would imply a much enhanced branching ratio.
Glueball production from this rare semi-inclusive 
$B$ decay can be probed at the LHCb and  Belle II to narrow down 
the allowed parameter space.  Similar branching ratio is expected
for the pseudoscalar glueball. We also briefly comment on the case of 
vector and tensor glueballs.

\end{abstract}

\maketitle


\section{Introduction}

Despite the fact that a glueball state has not been confirmed experimentally,
its existence is an unavoidable and yet uncannily prediction of QCD.
Glueball states may have different Lorentz structures in general, such as a scalar, 
a pseudoscalar or even a tensor with either normal or exotic $J^{\rm PC}$ assignment.
The prediction for the glueball masses is, however, a difficult task.
Theoretical calculations indicate that the lowest lying glueball
state is a scalar $0^{++}$ state with a mass in the range of 1.6 to 2 GeV. 
Previous quenched lattice QCD calculation gave such a glueball mass $m_G$ equals $1.71\pm 0.05 \pm
0.08$ GeV \cite{lattice-quenched}. Recent result from unquenched lattice QCD calculation \cite{Gregory:2012hu} 
gives a result of $1.795\pm 0.060$ GeV for this state.
These results support that the scalar
meson $f_0(1710)$ to be a glueball. Phenomenologically,
$f_0(1710)$ could be an impure glueball since it may be
contaminated by possible mixings with the colorless quark-antiquark states
that have total isospin zero
\cite{lee-weingarten,burakovsky-page,giacosa-etal,close-zhao,he-li-liu-zeng,cheng-chua-liu,fariborz}. These mixing
effects can be either small \cite{giacosa-etal,close-zhao,he-li-liu-zeng, fariborz} or
large \cite{lee-weingarten,burakovsky-page,giacosa-etal,cheng-chua-liu},
depend largely on the mixing schemes
one chose to do the fits and these may complicate the analysis. 

Quenched lattice calculation also showed that the lowest lying $0^{-+}$ state for a pure pseudoscalar glueball may have a higher mass around $2.560 \pm 0.035 \pm 0.120$ GeV \cite{lattice-quenched}. 
QCD sum-rule approach \cite{Narison:1996fm,Forkel:2003mk,He:1997xk,Gabadadze:1997zc} 
also predicted higher than 1.8 GeV for pseudoscalar glueball mass.
All these results do not favored the earlier speculation of the $\eta(1405)$ being a pseudoscalar glueball.
$\eta(1405)$ is indeed a perfect candidate for a $0^{-+}$ glueball since it is copiously produced 
from the radiative decay of $J/\psi$ and not seen in the $\gamma\gamma$ mode. 
Recently it has been shown that when mixing effects and related data for the transition matrix elements of 
anomalous axial-vector current between vacuum and the three states 
$\vert \eta\rangle$, $\vert \eta^\prime \rangle$ and $\vert \eta\left(1405 \right) \rangle$ 
are taken into account, the physical pseudoscalar glueball can be as light as $1.4 \pm 0.1$ GeV \cite{Cheng-Li-Liu}. 
Clearly, a full unquenched lattice calculation with the fermion determinant 
included must be performed in order to settle this issue satisfactory. 
Indeed, it has been argued some time ago \cite{Gabadadze:1997zc} that due to the dynamical fermion effects
the full QCD prediction for the pseudoscalar glueball mass
will be substantially departure from its quenched approximation. 
However, the mass for the lowest lying $0^{-+}$ state is still missing in the 
latest unquenched lattice results \cite{Gregory:2012hu}.

Understanding glueball dynamics is therefore important though very challenging. 
In this work we will study scalar or pseudoscalar glueball production from 
rare semi-inclusive $B$ meson decay.
The main purpose of this paper is to refocus the attention of experimentalists for 
undertaking the proposed searches. 
From the experimental findings we can learn a lot. The quantitative numerical predictions are 
bound to have significant theoretical uncertainties; nevertheless experimental efforts will be useful.

For a recent review for the phenomenology of scalar and pseudoscalar glueballs, we refer to Ref.~\cite{haiyang-glueballs}, and
for a recent summary of glueballs on the lattice, see Ref.~\cite{Lucini:2014paa}.

We lay out this paper as the following.
In section II, we discuss the effective vertex of $b \to s g^*$ induced by the gluonic penguin 
and the effective couplings between scalar and pseudoscalar glueballs with the gluons.
In section III, we discuss the interaction of scalar glueball  
with the light pseudoscalar meson octet using the chiral Lagrangian technique. 
The pseudoscalar glueball case will be briefly mentioned as well.
In section IV, the rates of the semi-inclusive $B$ meson decay into scalar and pseudoscalar glueballs are presented.
In section V, mixing effects for both scalar and pseudoscalar glueballs will be considered.
We summarize our results in section VI.


\section{Effective Couplings}

\begin{figure}[hbt]
\begin{center}
\includegraphics[width=9.cm]{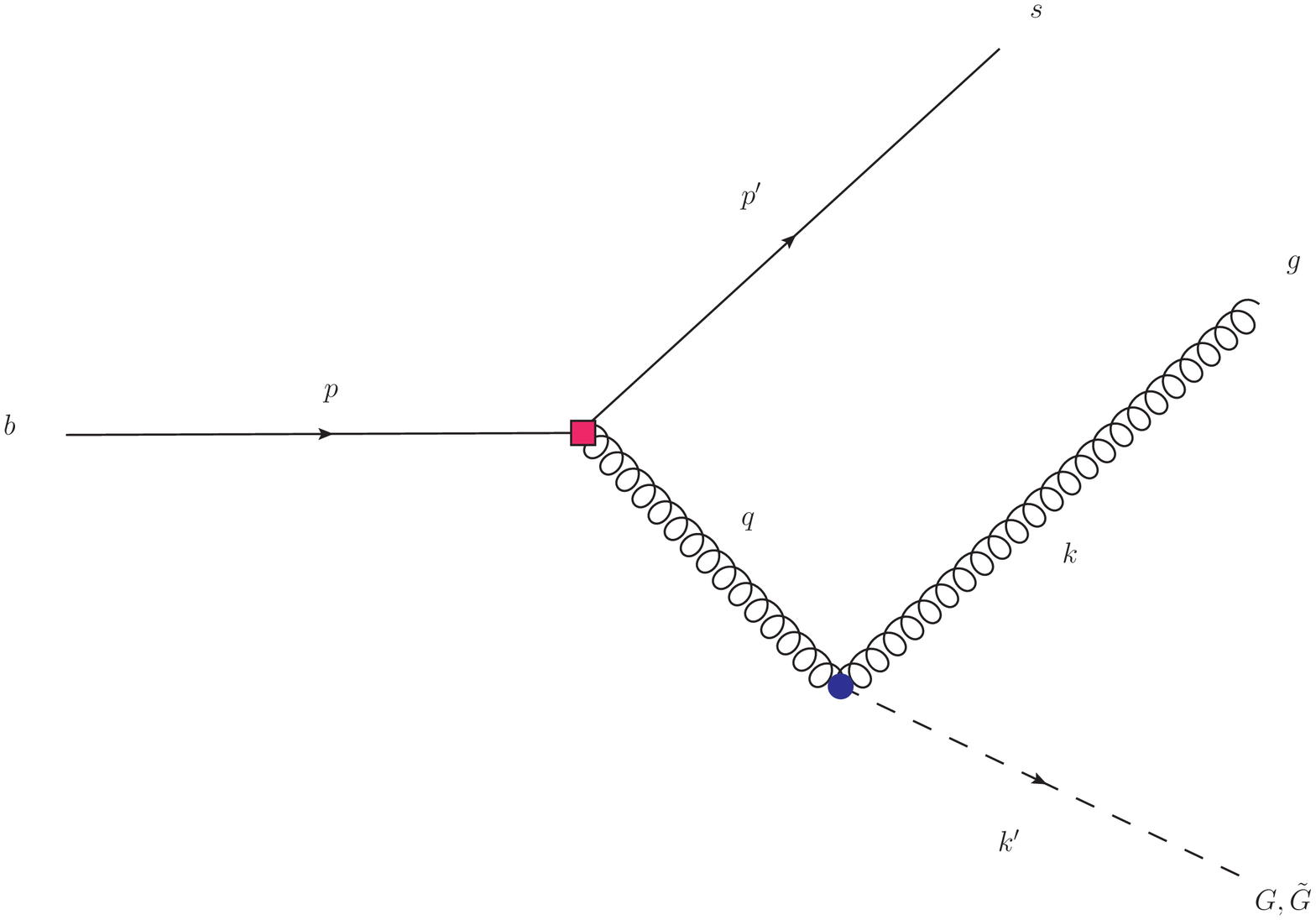}
\end{center}
\caption{The leading diagram for $B\to \mbox{Glueball} + X_s$.} 
\label{Feynman-1}
\end{figure}

Since the leading Fock space of a glueball $G$ is made up of two
gluons, production of glueball is therefore most efficient at a
gluonic rich environment like $J/\psi$ or $\Upsilon$  $\rightarrow
(gg)\gamma \rightarrow G\gamma$ \cite{shifman,jpsi}. Direct
glueball production is also possible at the $e^+e^-$ \cite{brodsky}
and hadron colliders.
In this work, we show that
$B\to G + X_s$ decay also provides an interesting mechanism to produce and to detect
a glueball. The
leading contribution for this process is shown in Fig.~[\ref{Feynman-1}], 
where the squared vertex refers to the gluonic
penguin interaction and the round vertex stands for an effective
coupling between a glueball and the gluons.
The gluonic penguin $b\to s g^*$ has been studied extensively in
the literature and was used in the context for inclusive
decay $b \to s g \eta'$ \cite{hou-tseng}. The effective
interaction for $b\to s g^*$ with next-to-leading QCD correction
can be written as \cite{he-lin}
\begin{eqnarray}
\Gamma_{\mu a} =- \frac{G_F}{\sqrt{2}}\frac{g_s}{4 \pi^2}
V_{ts}^*V_{tb} \bar s(p') \, [ \Delta F_1(q^2 \gamma_\mu - q_\mu \not\!
q) L - i m_b F_2  \sigma_{\mu\nu} q^\nu R ] \, T^a \, b(p),
\label{penguin}
\end{eqnarray}
where $\Delta F_1 = 4\pi(C_4(q, \mu) + C_6(q,
\mu))/\alpha_s(\mu)$ and $F_2 = -2 C_8(\mu)$ with $C_i(q,\mu)$  $(i=4,6, \, {\rm and} \, 8$) the
Wilson's coefficients of the corresponding operators in the
$\Delta B = 1$ effective weak Hamiltonian, $q = p - p'=k+k'$, and
$T^a$ is the generator for the color group. We will use the
next-to-leading order numerical values of $\Delta F_1$ and $F_2$ \cite{he-lin}. The top
quark contribution gives $\Delta F^{\rm top}_1 = -4.86$ and $F^{\rm top}_2 = +0.288$
at $\mu = 5$ GeV; whereas the charm quark contribution involves a $q^2$ dependence
through $C_4^{\rm charm}(q,\mu) = C^{\rm charm}_6(q,\mu) = P^{\rm charm}_s(q,\mu)$ with
\begin{eqnarray}
P^{\rm charm}_s(q,\mu) & = & {\alpha_s(\mu) \over 8 \pi} C_2(\mu) \left (
{10\over 9} + Q(q,m_c,\mu) \right) \; ,
\end{eqnarray}
and
\begin{eqnarray}
Q(q,m,\mu) & = &  4\int^1_0 dx \, x\,(1-x) \ln \left[m^2 - x(1-x)q^2\over \mu^2 \right]
 \; \, \nonumber \\
& = &
\frac{2\left[ 3q^2\ln (m^2/\mu^2)-12m^2-5q^2 \right]}{9q^2} \\
&& \; + \; 
\frac{4(2m^2+q^2)\sqrt{4m^2-q^2}}{3\sqrt[3]{q^2}}{\rm arctan}\sqrt{\frac{q^2}{4m^2-q^2}} \; .
\nonumber
\end{eqnarray}
Here $m_c = $ 1.4 GeV is the charm quark mass and $C_2(\mu = 5\,\mbox{GeV}) = 1.150$.

The following effective coupling between a scalar glueball and two
gluons can be parametrized as \cite{Carlson:1980kh,cornwall-soni,chanowitz}
\begin{eqnarray}
{\cal L}_0 = f {\cal G} G^a_{\mu\nu}G^{a\mu\nu},\label{chano}
\end{eqnarray}
where ${\cal G}$ is the interpolating field for the glueball $G$,
$G^a_{\mu\nu}$ is the gluon field strength, and $f$ is an unknown
coupling constant.  When combined with the interaction vertex in Eq.~(\ref{penguin}), 
a scalar glueball can be produced in 
rare $b$ quark decay, $b \to s g G$ (Fig.~(\ref{Feynman-1})). 
At the hadronic level, this leads to the rare semi-inclusive decay $B \to G + X_s$.
The above effective coupling has also been used for the study of exclusive decays of
$B^\pm \to f_0(1710) + K^{(*)\pm}$ \cite{Chen:2007uq}.

Analogously one can also study inclusive $B$ decays into a
pseudoscalar glueball $\tilde G$, whose leading effective coupling to the gluon can be
parameterized as \cite{Carlson:1980kh,cornwall-soni,chanowitz},
\begin{eqnarray}
{\cal L}_0^\prime = \tilde f \tilde {\cal G} \tilde G^a_{\mu\nu}
G^{a\mu\nu} \; \;\;\; {\rm with} \;\;\; \; \tilde G^a_{\mu\nu} =
{1 \over 2}\epsilon_{\mu\nu\alpha\beta} G^{a\alpha\beta}  \; ,
\end{eqnarray}
where $\tilde{\cal G}$ denoting the interpolating field for the
pseudoscalar glueball and $\tilde f$ is an unknown coupling. When combined with the interaction 
vertex in Eq.~(\ref{penguin}), 
a pseudoscalar glueball can be produced in 
rare $b$ quark decay, $b \to s g \tilde G$ (Fig.~(\ref{Feynman-1})). At the hadronic level, this leads to the rare 
semi-inclusive decay $B \to \tilde G + X_s$.


\section{Decay Rates}

If glueballs are composed of only gluons, one can easily obtain the branching ratios for $B\to G +  X_s$.
With the two effective couplings given in Eqs.~(\ref{penguin}) and (\ref{chano}),
the following decay
rate for $b(p) \rightarrow s(p') g (k) G (k')$ (Fig.~(\ref{Feynman-1})) can be obtained readily
\begin{eqnarray}
\Gamma_{b \rightarrow sgG}
& = & \left( {N_c^2 - 1 \over 4 N_c} \right) {G_F^2 m_b^5 \vert V^*_{ts} V_{tb} \vert^2 \over 2^5 \pi^3}
\left( {g_s \over 4 \pi^2} \right)^2 (m_b f)^2 \nonumber\\
& \;\; & \times \int_0^{(1-\sqrt{r})^2} dx \int_{y_-}^{y_+} dy \left\{
|\Delta F_1|^2 c_0 + {\rm\bf Re}(\Delta F_1 F^*_2) {c_1\over y} +
|F_2|^2 {c_2\over y^2} \right\} \; ,
\label{dis}
\end{eqnarray}
with
\begin{eqnarray}
y_{\pm} = {1 \over 2} \left[ (1-x+r) \pm \sqrt{(1-x+r)^2 - 4 r} \right] \; .
\end{eqnarray}
In Eq.~(\ref{dis}), $N_c$ is the number of color
and $c_{0,1,2}$  are given by
\begin{eqnarray}
c_0 & = & \frac{1}{2} [-2x^2 y + (1-y)(y-r)(2x+y-r)] \; ,\nonumber\\
c_1 & = & (1-y)(y-r)^2 \; ,\nonumber\\
c_2 & = & \frac{1}{2} [2x^2y^2 -(1-y)(y-r)(2xy-y+r)] \; ,
\end{eqnarray}
with $x = (p' + k)^2/m^2_b$,  $y=(k+k')^2/m^2_b$, and
$r=m^2_G/m^2_b$.

The decay rate for pseudoscalar glueball case $B\to \tilde G + X_s$ can be deduced
from the scalar glueball one by replacing the coupling $f$ with $\tilde f $ and
the mass $m_G$ with the pseudoscalar glueball mass $m_{\tilde G}$ in Eq.~(\ref{dis}). 
This also reproduces previous result for the $x-y$ distribution 
obtained in Ref.~\cite{hou-tseng} for a similar process $b \rightarrow sg\eta^\prime$. 
Thus it is not possible to differentiate the produced glueball is a scalar or pseudoscalar 
from the $x-y$ distribution of this semi-inclusive $B$ decay.


\section{Interaction of glueballs with light mesons}

The production rate of a scalar or a pseudoscalar glueball from $B$ decays depend on the strength of the couplings $f$ 
or $\tilde f$ respectively. 
We first discuss how to use the chiral Lagrangian methods to incorporate the interaction 
for a scalar glueball with the other light mesons and then employ experimental data to 
determine $f$.

The form of effective coupling of a scalar glueball with two gluons, ${\cal L}_0 = f {\cal G} G^a_{\mu\nu}G^{a\mu\nu}$, suggests that glueball couples to the QCD trace anomaly.
As briefly mentioned in \cite{chm},
the interaction of a scalar glueball with light hadrons through the
trace anomaly can be
formulated systematically by using techniques of chiral Lagrangian.
The kinetic energy and the symmetry breaking mass terms for the light
pseudoscalar mesons are given by \cite{donoghue}
\begin{eqnarray}
{\cal L}_\chi = {f^2_\pi \over 8} [ {\rm Tr}(\partial^\mu \Sigma
\partial_\mu \Sigma^\dagger) +  {\rm Tr}(\xi^\dagger \chi \xi^\dagger + \xi \chi
\xi)],
\end{eqnarray}
where  $f_\pi =   132$ MeV being the pion decay constant,
$\xi^2 =\Sigma = \exp (2i\Pi/f_\pi)$ with $\Pi$ the $SU(3)$
pseudoscalar meson octet,
\begin{eqnarray}
\Pi = \left ( \begin{array}{cccc}
{1 \over \sqrt{2}}\pi^0+{1 \over \sqrt{6}}\eta  &\pi^+ &K^+\\
\pi^-&- {1 \over \sqrt{2}}\pi^0 +{1 \over \sqrt{6}}\eta&K^0\\
K^-&{\overline K}^0 &-{2 \over \sqrt{6}} \eta \end{array}\right ) \;,
\end{eqnarray}
and
\begin{eqnarray}
\chi= 2B_0 \, {\rm diag}(\hat m, \hat m, m_s) = {\rm diag}(m^2_\pi, m^2_\pi,
 2m^2_K - m^2_\pi)
 \end{eqnarray}
with $B_0 = 2031$ MeV.
Here we have neglected the isospin breaking effects due to small mass
difference between the light $u$ and $d$ quarks and used $\hat m = m_u = m_d$.
The QCD trace anomaly is well known and given by \cite{donoghue}
\begin{eqnarray}
\Theta^\mu_\mu
& = & - \frac{b\alpha_s}{8\pi} G^a_{\mu\nu}G^{a\mu\nu} + \sum_q m_q \bar q q \; ,
\end{eqnarray}
where $b = 11 - 2 n_f / 3$ is the QCD one-loop beta function with $n_f = 3$ being the number of light quarks.
Treating the effective interaction (\ref{chano}) as a perturbation to the energy
momentum stress tensor, one would then modify $\Theta^{\mu}_{\mu}$ to be
\begin{eqnarray}
-\frac{b\alpha_s}{8\pi} G^a_{\mu\nu}G^{a\mu\nu} \left(1+ f \frac{8\pi }{ b \alpha_s} {\cal G} \right) +
\sum_q \left( m_q +f_q {\cal G} \right) \bar q q  \; ,
\end{eqnarray}
where $f_q = f\alpha_s
m_q(16\pi\sqrt{2}/3\beta)\ln\left[(1+\beta)/(1-\beta)\right]$ being the one-loop induced $G q \bar q$
coupling \cite{chanowitz} with $\beta =
(1-4m^2_q/m_G^2)^{1/2}$. Note that $f_q$ is proportional to $f$.
The corresponding chiral Lagrangian is thus modified accordingly
\begin{eqnarray}
{\cal L}_\chi &=&{1\over 8} f^2_\pi \left(1+  f {8\pi\over b \alpha_s} {\cal G} \right)
{\rm Tr} \left[ \partial^\mu \Sigma \partial_\mu \Sigma^\dagger \right] \nonumber\\
&+&{1\over 8} f^2_\pi \, {\rm Tr} \left[ \xi^\dagger (\chi+2B_0 f_\chi
{\cal G})\xi^\dagger + \xi (\chi + 2 B_0 f_\chi {\cal G}) \xi \right],
\label{trace}
\end{eqnarray}
where $f_\chi = {\rm diag}(f_u, f_d, f_s)$. 

Using the above chiral Lagrangian, one can calculate
the decay rates for $G \rightarrow \pi^+\pi^-,\; \pi^0\pi^0$, $K^+K^-,\; K^0{\overline K}^0$, and $\eta^0\eta^0$.
At present there are some data on the ratios of decay width, 
$\Gamma(\pi\pi)/\Gamma(K\bar K) \sim 0.41^{+0.11}_{-0.17}$ and 
$\Gamma(\eta\eta)/\Gamma(K\bar K) \sim 0.48\pm0.15$ \cite{pdg}.
Theoretically, we obtain $\Gamma(\pi\pi)/\Gamma(K\bar K) \sim 0.36$ and 
$\Gamma(\eta\eta)/\Gamma(K\bar K) \sim 0.30$ which agree with data at about  $1 \sigma$ level. 
With the above chiral Lagragian Eq.~(\ref{trace}), one can also obtain information about the size of the coupling $f$ if 
combined with the branching ratio ${\rm Br}(f_0(1710) \rightarrow K
\overline K) = 0.36\pm0.12$ \cite{brr} and the total width $\Gamma_{f_0(1710)} = 137\pm
8$ MeV for $f_0(1710)$ \cite{pdg}. Using a value of the
strong coupling constant $\alpha_s = 0.35$ extracted from the experimental data
of $\tau$ decay, we estimate the central value of the unknown coupling 
$f =0.07$ ${\rm GeV}^{-1}$. 

In our estimation of $f$ given above, we have extrapolated low
energy theorems to the glueball mass scale 
which is higher than $\Lambda_{\rm QCD}$.
The use of chiral Lagrangian is quite questionable.
One might overestimate
the hadronic matrix elements in due course. This implies that the
extracted value of $f$ would be too small. 

An alternative approach is to use perturbative QCD.
Indeed, this calculation was carried out some time ago for the decay $G \to K\overline K$ in \cite{chm}.
This approach can also give an estimate
of the amplitude. The problem facing this approach is that the
energy scale may not be high enough to have the perturbative QCD contribution
to dominate. 
Using the asymptotic light-cone wave functions, we
find that the branching ratios for $G \to K\overline K$ and $G \to \pi\pi$ decays
are proportional to $f_K^2$ and $f_\pi^2$, respectively. This leads to the ratio of
$\Gamma(\pi\pi)/\Gamma(K\bar K)$ to be about 0.48 in agreement with data within error bar.
Fitting branching ratio for $G\to K\overline K$, we find that the value of $f$ would be about 20 times 
larger than that obtained above using the chiral  approach. Incidentally, the value
of $f$ derived from the chiral Lagrangian is within a factor of 2
compared with that estimated just by using the free quark
decay rate of $G \rightarrow s \bar s$. 

In our later calculations, we will use the conservative value
$f=0.07$ ${\rm GeV}^{-1}$ determined using the
chiral Lagrangian given in Eq.~(\ref{trace}). This will give the
most conservative estimate for the branching ratio since the
chiral approach gives the smallest $f$.
Clearly, a rigourous determination of $f$ is on the lattice.
But we do not anticipate to have a lattice evaluation of its value any time soon.
We have to contend with its rough estimation from all the theoretical tools that are available
so far. Despite the fact one can not have a precise determination of the effective coupling $f$, 
we believe that glueball production from the proposed rare semi-inclusive $B$ decay is 
a very interesting mechanism which is worth searching for experimentally.

\begin{figure}[hbt]
\begin{center}
\includegraphics[width=10cm]{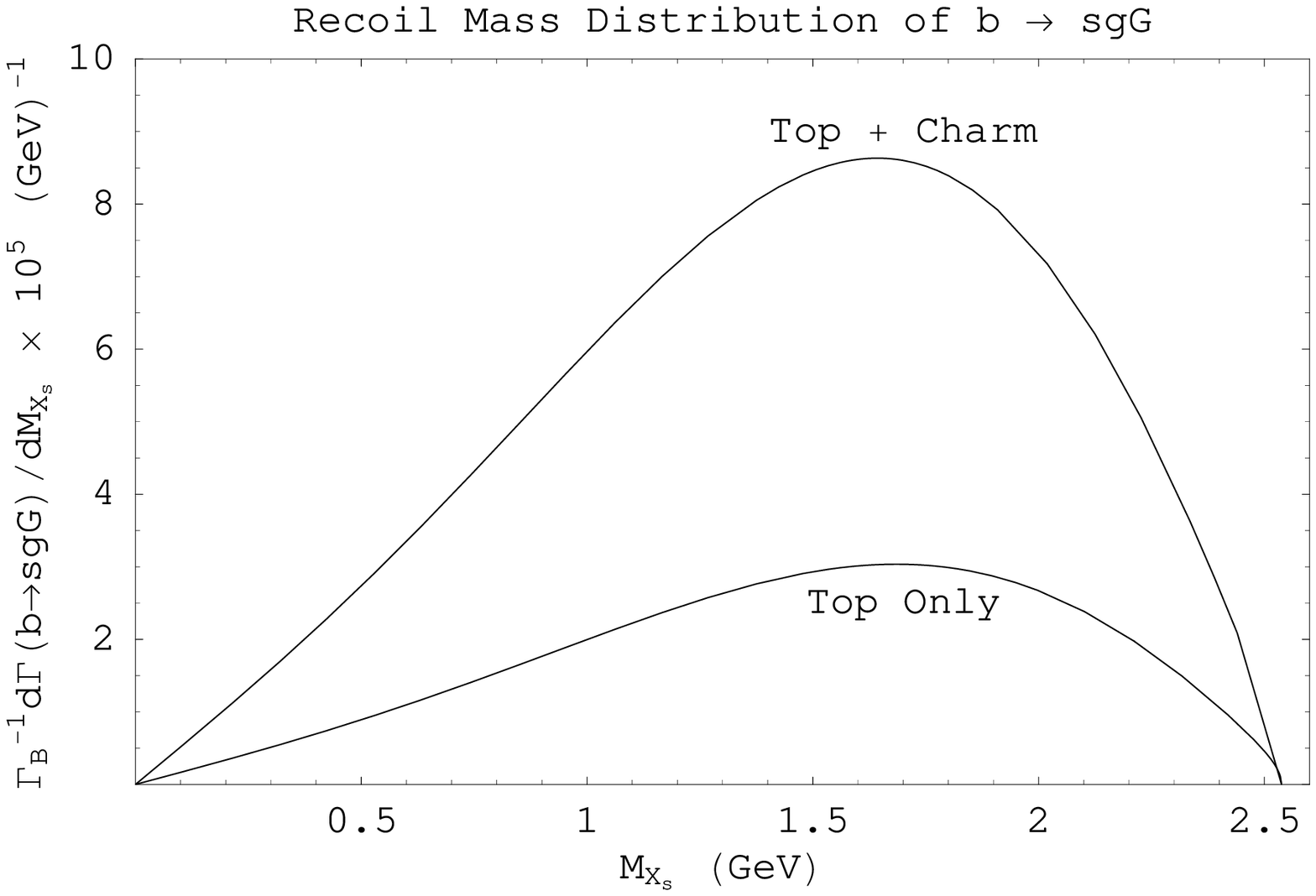}
\end{center}
\caption{$d{\rm Br}(B\to G + X_s)/dM_{X_s}$ in unit of $10^{-5}$. In
the figure the $b$ quark mass $m_b$ is taken to be 4.248 GeV,
$\alpha_s = 0.21$ at the $b$ quark scale, $\tau_B = 1.674 \times
10^{-12}$ s and $f = 0.07$ GeV$^{-1}$.} \label{Feynman-2}
\end{figure}

Using the value of $f$ determined above from the chiral
Lagrangian, we find the branching ratio for $b\to s  g G$ to
be $4.5\times 10^{-5}(f \cdot \mbox{GeV}/0.07)^2$ with just the leading
top penguin contribution to $\Delta F_1$ is taken into account. The correction from the
charm penguin is nevertheless substantial and should not be neglected. Inclusion of both top and charm
penguins gives rise to an enhancement about a factor of 3 in the branching ratio
${\rm Br}(b \to s gG) \approx
1.3\times10^{-4}(f \cdot \mbox{GeV}/0.07)^2$. Since $f_0(1710)$ has a
large branching ratio into $K\overline K$, the signal of scalar
glueball can be identified by looking at the secondary $K
\overline K$ invariant mass. The recoil mass spectrum of $X_s$ can
also be used to extract information. The distribution of $d{\rm
Br}(b \to s g G)/dM_{X_s}$ as a function of the recoil mass of
$X_s$ is plotted in Fig.~(\ref{Feynman-2}) where $M_{X_s}^2 = (p' + k)^2 = m_b^2 x $. 

Due to parity conservation, a pseudoscalar glueball will not decay into two pseudoscalar mesons.
Instead it will decay into one scalar and one pseudoscalar mesons or three pseudoscalar mesons.
If $\eta(1405)$ is indeed the lowest lying pseudoscalar glueball, kinematics allows it to decay into 
$f_0(980) \eta$, $K \overline K \pi$, $\eta \pi \pi$ and so on,  
as these modes were indeed reported as {\it seen} by Particle Data Group \cite{pdg}.
The interaction of pseudoscalar glueball with the light scalar and pseudoscalar meson octets
has been studied in Ref.~\cite{Eshraim:2012jv} using the chiral Lagrangian techniques 
for the resonance $X(2370)$ identified as a pseudoscalar glueball. 
One could apply the chiral Lagrangian given in Ref.~\cite{Eshraim:2012jv} to the lighter state $\eta(1405)$.
However the available phase space for $\eta(1405)$ to decay is much smaller than $X(2370)$.
We will not pursuit that further here.


\section{Mixing Effects}

As is well known, glueballs can mix with other pure $q \bar q$ states with the 
same quantum numbers, which had complicated the identification of glueballs experimentally.
For these mixing effects, we will adopt the mixing models of Ref.~\cite{cheng-chua-liu}
for the scalar glueballs and Ref.~\cite{Cheng-Li-Liu} for the pseudoscalar glueballs.
For the scalar case, we have \cite{cheng-chua-liu}
\begin{equation}
\label{scalarmixingmatrix}
\left( 
\begin{array}{c}
\vert f_0(1370) \rangle \\
\vert f_0(1500) \rangle \\
\vert f_0(1710) \rangle
\end{array}
\right) = \left( 
\begin{array}{ccc}
0.78 & 0.51 & -0.36 \\
-0.54 & 0.84 & 0.03 \\
0.32 & 0.18 & 0.93
\end{array}
\right) \cdot 
\left(
\begin{array}{c}
\vert N \rangle \\
\vert S \rangle \\
\vert G \rangle
\end{array}
\right) \; .
\end{equation}
Inverting the above relation, we obtain
\begin{equation}
\label{G2fs}
\vert G \rangle = -0.36 \vert f_0(1370) \rangle + 0.03 \vert f_0(1500) \rangle + 0.93 \vert f_0(1710) \rangle \; .
\end{equation}
Thus the above mixing induces $B\to f_0(1710)X_s$,  $B\to f_0(1370)X_s$ and $B\to f_0(1500) X_s$ with 
the following relative branching ratios
\begin{eqnarray}
\label{BRscalars}
{\rm Br}(f_0(1710)):{\rm Br}(f_0(1370)) :{\rm Br}(f_0(1500)) = (0.93)^2: (-0.36)^2 : (0.03)^2   
= 1 : 0.15 : 0.001\;,\nonumber\\
\end{eqnarray}
up to the corrections from the mass differences among the three states.

Similarly, for the pseudoscalar case, we have \cite{Cheng-Li-Liu}
\begin{equation}
\label{pseudoscalarmixingmatrix}
\left( 
\begin{array}{c}
\vert \eta \rangle \\
\vert \eta' \rangle \\
\vert \eta(1405) \rangle
\end{array}
\right) = \left( 
\begin{array}{ccc}
0.749 & -0.657 & 0.085 \\
0.600 & 0.728 & 0.331 \\
-0.279 & -0.197 & 0.940
\end{array}
\right) \cdot 
\left(
\begin{array}{c}
\vert \eta_8 \rangle \\
\vert \eta_1 \rangle \\
\vert {\tilde G} \rangle
\end{array}
\right) \; ,
\end{equation}
with
\begin{equation}
\label{PSG2fs}
\vert \tilde G \rangle = 0.085 \vert \eta \rangle + 0.331 \vert \eta' \rangle + 0.94 \vert \eta (1405) \rangle \; .
\end{equation}
Hence one induces $B\to \eta (1405) X_s $, $B\to \eta' X_s $ and $B\to \eta X_s$ with 
following relative branching ratios
\begin{eqnarray}
\label{BRpseudoscalars}
{\rm Br}(\eta (1405)):{\rm Br}(\eta'):{\rm Br}(\eta) = (0.94)^2 : (0.331)^2 : (0.085)^2 
= 1 : 0.12 : 0.008\;,\nonumber\\
\end{eqnarray}
up to the corrections from the mass differences among the three states.

We note that the above relative branching ratios were derived under the assumption that 
$B \to \left(G/\tilde G\right) + X_s$ arises purely from the three-body process 
$b \to s g \left( G/\tilde G \right)$ with the glueball mixing effects taken into account.
There may be two-body process $B \to \left(G / \tilde G \right) + X_s$ via 
$b\to s \left( G / \tilde G\right)$ from four-quark operators in the effective Hamiltonian 
for the $G /\tilde G$ production from $B$ decay. These contributions will modify the 
above predictions of the relative branching ratios. 
Being a two-body decay type, the energy of $G/\tilde G$ 
is energetic, given by $E^{\rm 2-body}_{G /\tilde G} = (m^2_B + m^2_{G / \tilde G} - m_{X_s}^2)/2m_B$ 
with $m_B = 5279.50 \pm 0.30$ MeV and $m_{X_s}$ typically equals 1 GeV or so. 
On the other hand, for the $G/\tilde G$ production induced by $b\to s g  \left( G/\tilde G \right)$, the
energy of $G/\tilde G$ is more spread out. One therefore expects with a suitable cut on the glueball energy 
the two-body contributions can be removed, leaving the softer $G/\tilde G$ production presumably coming from 
the mechanism we suggested in this paper to satisfy the relative branching ratios given in this section.

To implement the cut, recall that the energy of the glueball $E_{G/\tilde G}$ is related to the rescaled variable 
$x = (p'+k)^2/m_b^2$ according to $x = 1+r-2 E_{G/\tilde G}/m_b$ where 
$r = m_{G / \tilde G}^2 / m_b^2$. 
With $E_{G/\tilde G}^{\rm max} = m_b (1 + r)/2$, $x_{\rm min} = 0$. 
Thus to make a cut on the maximum of $E_{G/\tilde G}$ corresponds to a cut on 
the minimum (lower limit in the integration) of $x$. In the two-body decay, the energy of the
glueball is fixed at $E^{\rm 2-body}_{G/\tilde G} = m_b (1 + r  - \tilde r)/2$ with $\tilde r = m_{X_s}^2 /m_b^2$. 
Since $m_{X_s}$ is of order 1 GeV, we will impose a cut on the minimum of $x$ as 
in the form $x^{\rm min} = \delta \cdot \tilde r$
by varying $\delta$ in a range from 1 to 3. The effects of these cuts on the branching ratios of 
Br($B \to f_0(1710)X_s$)  and Br($B \to \eta (1405)X_s$) are shown in Table~\ref{BrTable}.
The impact of these cuts reduce the branching ratios by at most a factor of 3.

\begin{table} 
\scriptsize
\centering
\begin{tabular}{| c | c | c | c |  c | c | c |}
\hline
$\delta$ & $x^{\rm min} = \delta \cdot \tilde r$ & $E^{\rm max}_{G/\tilde G} = \frac{m_b}{2} \left( 1 + r - x^{\rm min} \right)$ (GeV) & \multicolumn{2}{|c|}{Br($B \to f_0(1710)X_s) \times 10^5$} & \multicolumn{2}{|c|}{Br($B \to \eta (1405)X_s) \times 10^5$}\\ \hline\hline
& & &Top Only & Top + Charm & Top Only & Top + Charm\\ 
\hline
0 & 0 & 2.468 / 2.356 & 4.5 & 12.9 & 7.1 & 19.4 \\
\hline
1 & 0.055 & 2.350 / 2.238 & 3.6 & 10.2 & 5.9 & 16.0 \\
1.2 & 0.066 & 2.326 / 2.215 & 3.4 & 9.6 & 5.7 & 15.2\\
1.4 & 0.077 & 2.303 / 2.191 & 3.2 & 9.0 & 5.4 & 14.5 \\
1.6 & 0.088 & 2.279 / 2.168 & 3.0 & 8.4 & 5.1 & 13.7 \\
1.8 & 0.099 & 2.256 / 2.144 & 2.8 & 7.8 & 4.9 & 13.0 \\
2 & 0.111 & 2.232 / 2.121 & 2.6 & 7.2 & 4.6 & 12.2 \\
2.5 & 0.138 & 2.174 / 2.062 & 2.1 & 5.8 & 4.0 & 10.4 \\ 
3 &  0.166 & 2.115 /  2.003 &  1.6 & 4.5  & 3.3 & 8.6 \\
\hline
\end{tabular}
\caption{\label{BrTable} 
Branching ratios Br($B \to f_0(1710)X_s$)  and Br($B \to \eta (1405)X_s$) as function of the maximum energy 
(or minimum $x$) cut of the glueballs. We take the nominal value of $f = \tilde f = 0.07$ GeV$^{-1}$. 
}
\end{table}


\section{Discussions}

Before closing, we would like to make the following comments:

\begin{itemize}

\item[(1)]

In 2005, BESII experiment \cite{bes2} has observed an enhanced decay in $J/\psi \to \gamma
\eta' \pi\pi$ with a peak around the invariant mass of
$\eta'\pi\pi$ at 1835 MeV. Proposal has been made to
interpret this state $X(1835)$ to be due to a pseudoscalar
glueball \cite{hhh}. Later BESIII experiment \cite{Ablikim:2010au} has further 
confirmed the observation of $X(1835)$ and its spin is consistent with 
expectation for a pseudoscalar.
Taking $X(1835)$ to be a pseudoscalar
glueball, we would obtain a branching ratio of $3.7\times
10^{-5}(\tilde f \cdot  \mbox{GeV} / 0.07)^2$ with top penguin
contribution only, and is enhanced to $1.1 \times 10^{-4}(\tilde
f \cdot \mbox{GeV}/0.07)^2$ if charm penguin is also included. If the
coupling $\tilde f$ is of the same order of magnitude as $f$, the
branching ratio for $B\to X(1835) + X_s$ is also sizable.

\item[(2)]

Lattice QCD calculations also predicted lowest lying spin 2 tensor states:
The $2^{++}$ and $2^{-+}$ masses are  
$2390 \pm 30 \pm 120$ MeV and $3040 \pm 40 \pm 150$ MeV respectively 
from the earlier quenched calculations \cite{lattice-quenched}, and
$2620 \pm 50$ MeV and $3460 \pm 320$ MeV respectively from 
recent unquenched results \cite{Gregory:2012hu}.
These heavier tensor states can also be produced from the semi-inclusive $B$ decay
but with much smaller available kinematic phase space compared with the scalar and pseudoscalar states. 
Leading operators describing the couplings between these tensor glueballs with the gluons are
\begin{equation}
{\cal G}^{\mu \nu} G^a_{\mu \alpha} G^{a \alpha}_ {\nu} \; \;\;\; {\rm and} \;\;\; \;
\tilde{{\cal G}}^{\mu \nu} G^a_{\mu \alpha} {\tilde G}^{a \alpha}_ {\nu} \; ,
\end{equation}
where ${\cal G}^{\mu\nu}$ and $\tilde{{\cal G}}^{\mu \nu}$ are interpolating fields 
for the $2^{++}$ and $2^{-+}$ glueball states respectively. 
Note that $\tilde{{\cal G}}^{\mu \nu}$
is not the dual of ${\cal G}^{\mu\nu}$ despite the $\,\tilde{}$.
It is straightforward to extend the present analysis to these tensor states.

\item[(3)]

Lattice calculations \cite{Gregory:2012hu} also predicted lowest lying spin 1 vector glueball states
which would necessarily decay into three gluons. We can also use effective operators 
describing interactions between the vector glueballs and gluons. 
We will leave them as exercises for our readers.

\end{itemize}

To conclude, we have studied semi-inclusive production of a scalar/pseudoscalar
glueball in rare $B$ decay through the gluonic penguin and effective glueball-gluon interactions. 
The branching ratio for the scalar glueball case is found to be of the order $10^{-4}$. 
$B$ decays into a pseudoscalar glueball through gluonic penguin is also expected to be sizable.
Note that for the scalar glueball case, we have used a conservative estimate of the effective coupling 
$f$. Observation of $B \to X_s f_0(1710)$ at a branching ratio of order $10^{-4}$
or larger will provide an strong indication that $f_0(1710)$ is
mainly a scalar glueball. 

Recall that we have now more than 600 millions of
$B\overline B$ accumulated at Belle and more than 300 millions at
B{\tiny A}B{\tiny AR}.
Enhancement by a factor of 50 is expected at the Super B Factory with a designed luminosity of
$8 \times 10^{35}$ cm$^{-2}$ s$^{-1}$ \cite{Aushev:2010bq}.
Moreover, at the 14 TeV LHCb, the $b \bar b$ cross section can be as large as 500 $\mu$b.
One would expect about $50 \times 10^{12}$ $b \bar b$ pair with an integrated luminosity of 
100 fb$^{-1}$ at the LHCb. 
Probing the branching ratio ${\rm Br}(b \to s g G/\tilde G)$ at the level of $10^{-4}$ is quite feasible. 
One should emphasis that both chiral Lagrangian and perturbative QCD approaches used to extract the
value of the effective coupling $f$ is not precise. Despite this shortcoming, the glueball production mechanism 
suggested in this work is an interesting one.
Measurement of this type may also help to pin down the energy scale where the 
chiral Lagrangian is valid for hadronic matrix element calculations.
We strongly urge our experimental colleagues  to rekindle their interests in search 
for the various glueball states; in particular those produced by the semi-inclusive decay of the $B$ meson 
proposed in this work.


\acknowledgments 
We like to thank K.~Cheung and J.~P.~Ma for useful discussions.
Our special thanks go to H.~Y.~Cheng for sharing his immense knowledge of glueball physics.
The work was supported in part by MOE Academic Excellent Program (Grant No: 102R891505), MOST and NCTS of ROC, and in part by NSFC(Grant No:11175115) and Shanghai Science and Technology Commission (Grant No: 11DZ2260700) of PRC. 

\vspace{1cm}


\end{document}